\documentclass{osa-article}
\journal{osajournal}
\articletype{Research Article}

\usepackage{float}
\usepackage{color, soul}

\usepackage{siunitx}

\begin{document}

\title{Optical back-action on the photothermal relaxation rate}

\author{Jinyong Ma, Giovanni Guccione, Ruvi Lecamwasam, Jiayi Qin, Geoff T.\ Campbell, Ben C.\ Buchler, Ping Koy Lam{*}}

\address{Centre for Quantum Computation and Communication Technology, Department of Quantum Science, Research School of Physics and Engineering, The Australian National University, Canberra ACT 2601, Australia}

\email{\authormark{*}ping.lam@anu.edu.au} 



\begin{abstract}
Photothermal effects can alter the response of an optical cavity, for example, by inducing self-locking behavior or unstable anomalies. The consequences of these effects are often regarded as parasitic and generally cause limited operational performance of the cavity. Despite their importance, however, photothermal parameters are usually hard to characterize precisely. In this work we use an optical cavity strongly coupled to photothermal effects to experimentally observe an optical back-action on the photothermal relaxation rate. This effect, reminiscent of the radiation-pressure-induced optical spring effect in cavity optomechanical systems, uses optical detuning as a fine control to change the photothermal relaxation process. The photothermal relaxation rate of the system can be accordingly modified by more than an order of magnitude. This approach offers an opportunity to obtain precise in-situ estimations of the parameters of the cavity, in a way that is compatible with a wide range of optical resonator platforms. Through this back-action effect we are able to determine the natural photothermal relaxation rate and the effective thermal conductivity of the cavity mirrors with unprecedented resolution.
\end{abstract}

\section{Introduction}

Light is a powerful tool in engineering the dynamics of the system with which it interacts. A well-known example is the optical spring \cite{BraginskyOptical1997, SheardObservation2004, MizrahiTwoslab2007,  SinghAllOpticalOptomechanicsOptical2010, Hossein-Zadehopticalspring2007, WiederheckerControlling2009, VogelOptically2003, kelley_observation_2015} and damping effects \cite{CohadonCoolingMirrorRadiation1999,kippenberg_analysis_2005, chan_cooling_opto, marquardt_quantum_2007,peterson_laser_2016, clark_sideband_2017} observed in optomechanical systems where a mechanical oscillator and an optical cavity are coupled via the radiation pressure force of light. An anti-restoring force and a viscous damping force induced by the radiation pressure force are exhibited in the red-detuned regime where one can exploit the optical damping effect to cool the mechanical oscillator to its ground state \cite{elste_quantum_2009, TeufelSideband2011, chan_cooling_opto}. A blue-detuned laser, on the other hand, produces an optical restoring force as well as an anti-damping force. The restoring force has led to applications in optical trapping and levitation \cite{singh_all-optical_2010, guccione_scattering-free_2013} while the anti-damping force has a critical role in the arousal of self-sustained oscillations \cite{carmon_temporal_2005-1, metzger_self-induced_2008} and chaotic behaviors \cite{carmon_chaotic_opto, sun_chaotic_2014, MaFormation2014} in the system. The natural relaxation rate of photothermal effects in an optical cavity also has a distinctive dependence on the detuning of the driving field. Even though it develops from a very different dynamical process, this phenomenon has similar properties to the radiation-pressure-induced optical spring. In particular, it can be employed to characterize photothermal parameters in a cavity with unprecedented resolution.
 
Characterizing the photothermal parameters and dynamics can reveal important information about the system. On the one hand, photothermal effects can impose a limit to state-of-the-art displacement measurements of high-sensitivity interferometers~\cite{BraginskyThermodynamical1999a, BraginskyThermorefractive2000,evans_thermo-optic_2008, ColeTenfoldreductionBrownian2013}, from small AFM-cantilever optical cavities~\cite{KlecknerHighFinesseOptoMechanical2006} to kilometre-scale gravitational-wave detectors like LIGO~\cite{AbbottObservation2016,BraginskyThermodynamical1999a, HarryThermalnoiseoptical2006}. In extreme applications, the shot noise of the absorbed light can set a fundamental limit on the sensitivity of the measurements~\cite{LudlowCompact2007, cerdonio_thermoelastic_2001} and the production of low-frequency squeezing \cite{GodaPhotothermal2005}. On the other hand, photothermal effects were found to be effective in suppressing the Brownian noise of a mechanical oscillator \cite{metzger_cavity_2004, HosseiniMultimodelasercooling2014} and may even cool a mechanical resonator close to its quantum ground state \cite{pinard_quantum_2008}.

Braginsky et al.~\cite{BraginskyThermodynamical1999a} were the first to advance a model for photothermal effects in interferometric systems. Their analysis approximated the targets as half-infinite mirrors and is valid in the so-called adiabatic limit, where the thermal diffusion length is shorter than the beam spot radius. In terms of dynamical photothermal effects, this approximation corresponds to the regime of frequencies higher than a critical cut-off frequency determined by the photothermal relaxation rate. Cerdonio et al. proposed a complete model valid over the full dynamical range~\cite{cerdonio_thermoelastic_2001}. This model was soon confirmed experimentally~\cite{de_rosa_experimental_2002} and extended to account for thin-film coatings and higher-frequency corrections~\cite{rosa_experimental_2006}. Other recent approaches~\cite{BlackEnhancedphotothermaldisplacement2004, farsi_photothermal_2012}, despite being successful at an absolute calibration of the photothermal parameters, often required pump-probe schemes and involved complex fitting models. The photothermal parameters reported by these investigations have a relative uncertainty on the order of $\approx$ 10-20\%.

In this work, we report for the first time the explicit dependence of the natural relaxation rate of photothermal effects on cavity detuning, in analogy to how the mechanical frequency of a mirror is modified by the radiation pressure's optical spring effect in optomechanical systems. The back-action of the cavity modifies the relaxation rate to be faster by more than an order of magnitude. This can be crucial in the evaluation as well as in the control of the photothermal response of any cavity-based systems and can be applied to build optical filters with tunable critical cut-off frequency. Awareness of this optical correction is also important when exploring the complex dynamics of a hybrid system, such as optomechanical cavities strongly coupled to photothermal effects~\cite{AbbottObservation2016, LecamwasamDynamics2020,MaObservationaccepted}.

As a demonstration, we apply this back-action induced correction to precisely calibrate the photothermal parameters of our system. Unlike previous characterizations, our scheme and model present three unique properties. Firstly, our scheme takes advantage of photothermal self-locking of the cavity and can be performed in-situ with only a single laser beam, an optical modulator, and two photodetectors, with no need for external feedback control. Secondly, our method is compatible not only with amplitude modulation but also with phase modulation of the cavity field, generalizing our scheme to a broader set of experiments. Finally, our proposal offers a more concise measurement with only two free parameters, allowing for precise fitting to experimental data. Experimentally, we show that the altered photothermal relaxation rate can be an order of magnitude larger than its natural value.  Additionally, the best fit of the cavity phase response gives us the photothermal relaxation rate of $16.2\pm0.2$ \SI{}{\hertz}, corresponding to the thermal conductivity of $1.182\pm0.016$ \SI{}{\watt/\kelvin.\meter}. The precision attained is about an order of magnitude better than previous works \cite{de_rosa_experimental_2002, farsi_photothermal_2012, BlackEnhancedphotothermaldisplacement2004}. 

\section{Modeling}
We consider an optical cavity driven by an intense laser field that heats the cavity mirrors and produces strong photothermal effects. We start with an empirical model that has been demonstrated~\cite{marino_chaotically_2011, KonthasingheSelfsustained2017} to describe the photothermal interactions. It is assumed that the change of cavity length increases (or decreases) exponentially as the cavity mirrors are heated up by the stationary intracavity optical field. Denoting by $q_\textrm{\rm th}$ the total change of the optical path length induced by the net photothermal effects and by $a$ the amplitude of the intracavity field, the dynamics of the photothermal interaction between the two can be modeled by the following two equations of motion, in the frame rotating at the driving laser frequency $\omega_l$:
\begin{eqnarray}
 \dot{q}_\textrm{\rm th} & = &-\gamma_{\rm th}(q_\textrm{\rm th}+\beta P_{c}) \label{eq:qth} ,\\
 \dot{a}& = & -\left[\kappa/2-i(\Delta+Gq_\textrm{\rm th})\right]a+\varepsilon_{\rm l}+\varepsilon_{\rm 0}\cos(\omega t+\text{\ensuremath{\varphi}}), \label{eq:a_back-action}
\end{eqnarray}
where $\gamma_{\rm th}$ is the photothermal relaxation rate, $\beta$ is the photothermal response coefficient, and $\kappa$ denotes the decay rate of the cavity. The term $-iGq_\textrm{\rm th}a$ in Eq.~(\ref{eq:a_back-action}) indicates the interaction between the cavity mode and photothermal displacement by the coupling constant $G=\omega_{\rm c}/L_{c}$ (where $\omega_{\rm c}$ is the cavity eigenfrequency and $L_{c}$ denotes the natural cavity length). The intracavity power is written explicitly as $P_{c}=\hbar\omega_{\rm c}\left\vert a\right\vert ^{2}/\tau_{\rm cav} = \hbar G c\left\vert a\right\vert ^{2}/2$ with $\tau_{\rm cav}=2L_{c}/c$ being the round-trip time of cavity photons. The cavity is driven at a detuning $\Delta$ from resonance, by an optical field of amplitude $\varepsilon_\textrm{l}$ which is sinusoidally modulated at a frequency $\omega$ and a phase $\varphi$ by a percentage $\varepsilon_0/\varepsilon_\textrm{l}\ll1$. We consider amplitude modulation here, however the methods works equally well with phase or frequency modulation (see Supplement 1). Note that the modulation is external and is applied for the purpose of characterizing the photothermal effects. 

Since the modulation depth is small, we can assume small deviations from the steady-state solutions and substitute the assumptions $q_\textrm{\rm th}=q_\textrm{\rm th}^{0}+\delta q_\textrm{\rm th}$ and $a=a_{0}+\delta a$ into Eqs.~(\ref{eq:qth})-(\ref{eq:a_back-action}) to obtain the steady states of the system:
\begin{eqnarray}
0 & = & q_\textrm{\rm th}^{0}+\alpha\left\vert a_{0}\right\vert ^{2}, \label{eq:steadyq}\\
0 & = & -\kappa a_{0}/2+i(\Delta a_{0}+Ga_{0}q_\textrm{\rm th}^{0})+\varepsilon_{\rm l}, \label{eq:steadya_back-action}
\end{eqnarray}
and the first-order dynamics:
\begin{eqnarray}
\delta \dot{q}_\textrm{\rm th} & = & -\gamma_{\rm th} \left[\delta q_\textrm{\rm th}+\alpha(a_{0}\delta a^{*}+a_{0}^{*}\delta a) \right] \label{eq:dq},\\
\delta \dot{a} & = & -\kappa\delta a/2+i\Delta_{\rm e}\delta a+iGa_{0}\delta q_\textrm{\rm th}\nonumber\\
&  & +\frac{\varepsilon_{\rm 0}}{2}e^{-i(\omega t+\varphi)}+\frac{\varepsilon_{\rm 0}}{2}e^{i(\omega t+\varphi)}, \label{eq:da}
\end{eqnarray}
where $\alpha=\hbar G c\beta/2$ and $\Delta_{\rm e} = \Delta+Gq_{\rm th}^{\rm 0}$. Combining Eqs. (\ref{eq:steadyq})-(\ref{eq:steadya_back-action}) gives us a cubic equation for $|a_0|^2$, implying that the system states can stay in a bistable regime at certain conditions~\cite{AnOptical1997, MaPhotothermally2020}. The cavity can be either self-locked or anti-locked depending on the scanning direction of the detuning. For example, consider the case where the photothermal effects manifest by expanding the medium of the cavity, thus lengthening the effective cavity length when power increases ($\beta>0$). A self-sustained equilibrium can be reached on the red-detuning side of resonance, where less intracavity power would shift the resonance back to lower frequencies and thus increase the power again. In cases where photothermal effects would instead decrease the effective length of the cavity ($\beta<0$), the self-locking mechanism is triggered on the opposite, blue-detuned side of resonance.

Assuming $\gamma_{\rm th}\ll\kappa$, we substitute the ansatz, $\delta q_\textrm{\rm th}  =  Q e^{-i\omega t}+Q^{*}e^{i\omega t}$ and $\delta a  =  A_{-}e^{-i\omega t}+A_{+}e^{i\omega t}$ , into Eq.~(\ref{eq:dq})-(\ref{eq:da}) and obtain the solution for the photothermal displacement which is expanded in the powers of $\frac{\gamma_{\rm th}}{\kappa/2}$ (see Supplement 1)
\begin{eqnarray}
    Q & = & \frac{\gamma_{\rm th}\alpha\varepsilon_{\rm l}\varepsilon_{\rm 0}e^{-i\varphi}}{(\Delta_{\rm e}^{2}+\kappa^{2}/4)}\frac{1}{i\omega-\gamma_{\rm th}+\zeta\gamma_{\rm th}}\times \label{eq:Qfull}\\
      &  & \left(1+\left(\frac{\gamma_{\rm th}}{\kappa/2}\right)\frac{i\omega}{\gamma_{\rm th}(\Delta_{\rm e}^2+\kappa^2/4)}\frac{i\omega-\gamma_{\rm th}-\gamma_{\rm th}\zeta}{i\omega-\gamma_{\rm th}+\gamma_{\rm th}\zeta}+\mathcal{O}\left(\frac{\gamma_{\rm th}}{\kappa/2}\right)^2\right), \nonumber
\end{eqnarray}
with
  \begin{eqnarray}
\zeta &=& \frac{\sigma\nu}{(\nu^{2}+1/4)^{2}}, \label{eq:zeta}
 \end{eqnarray}
where $\sigma = 2\alpha\varepsilon_{\rm l}^{2}G/\kappa^{3}$ and $\nu = \Delta_{\rm e}/\kappa$. The dimensionless parameter $\sigma$ is a compound coefficient, linearly proportional to the photothermal coefficient $\beta$ and inversely proportional to the cavity decay rate $\kappa$. It can be considered as an effective photothermal coefficient which intrinsic to the full cavity. In particular we note that characterizing the dependence of $\zeta$ on $\nu$ one can directly extrapolate the value of $\sigma$ and, therefore, the photothermal properties of the system. It is noted that $\zeta<0$ holds under self-locking conditions: if the photothermal coefficient is positive (i.e., $\beta>0$), the cavity can only be self-locked at red detuning (i.e., $\Delta_{\rm e}<0$). Similarly, $\Delta_{\rm e}$ is greater than zero when $\beta<0$. This indicates that $\sigma\nu$ and thus $\zeta$ is negative. The calibration of $\zeta$ which we will discuss later is subject to this property. We now look at the zero-order term of Eq. (\ref{eq:Qfull}) and define the term $\chi_{\rm th}(\omega) = (i\omega-\gamma_{\rm th}+\zeta\gamma_{\rm th})^{-1}$ to be the photothermal susceptibility, i.e. how the $x_{\rm th}$ degree of freedom responds to external optical field. The photothermal susceptibility is a single-pole function whose cut-off frequency corresponds, in the absence of photothermal-cavity interaction (i.e, $G = 0$ or $\zeta = 0$), to the natural photothermal relaxation rate $\gamma_{\rm th}$. When the photothermal effects are coupled to the intracavity optical field, however, the cut-off frequency of the photothermal responses is modified by 
\begin{eqnarray}
\delta \gamma_{\rm th} = -\zeta\gamma_{\rm th}. \label{eq:PTcorrection}
\end{eqnarray}
This optical correction phenomenon presents analogies to the optical spring effect driven by radiation pressure in optomechanical systems where the stiffness and damping rate of the mechanical oscillator are modified when interacting with the cavity field. Here, for the first time, we show that interaction with the cavity also modifies the response rate to photothermal excitations, $\gamma_{\rm th}$. We will also show that this effect can be exploited to allow precise characterization of the photothermal parameters of the cavity, with a relative precision that is an order of magnitude better than previous works~\cite{cerdonio_thermoelastic_2001, de_rosa_experimental_2002, farsi_photothermal_2012, BlackEnhancedphotothermaldisplacement2004} operating at the cavity resonance where $\delta\gamma_{\rm th}=0$.

\begin{figure}[htbp]
 \centering
\includegraphics[width=0.5\textwidth]{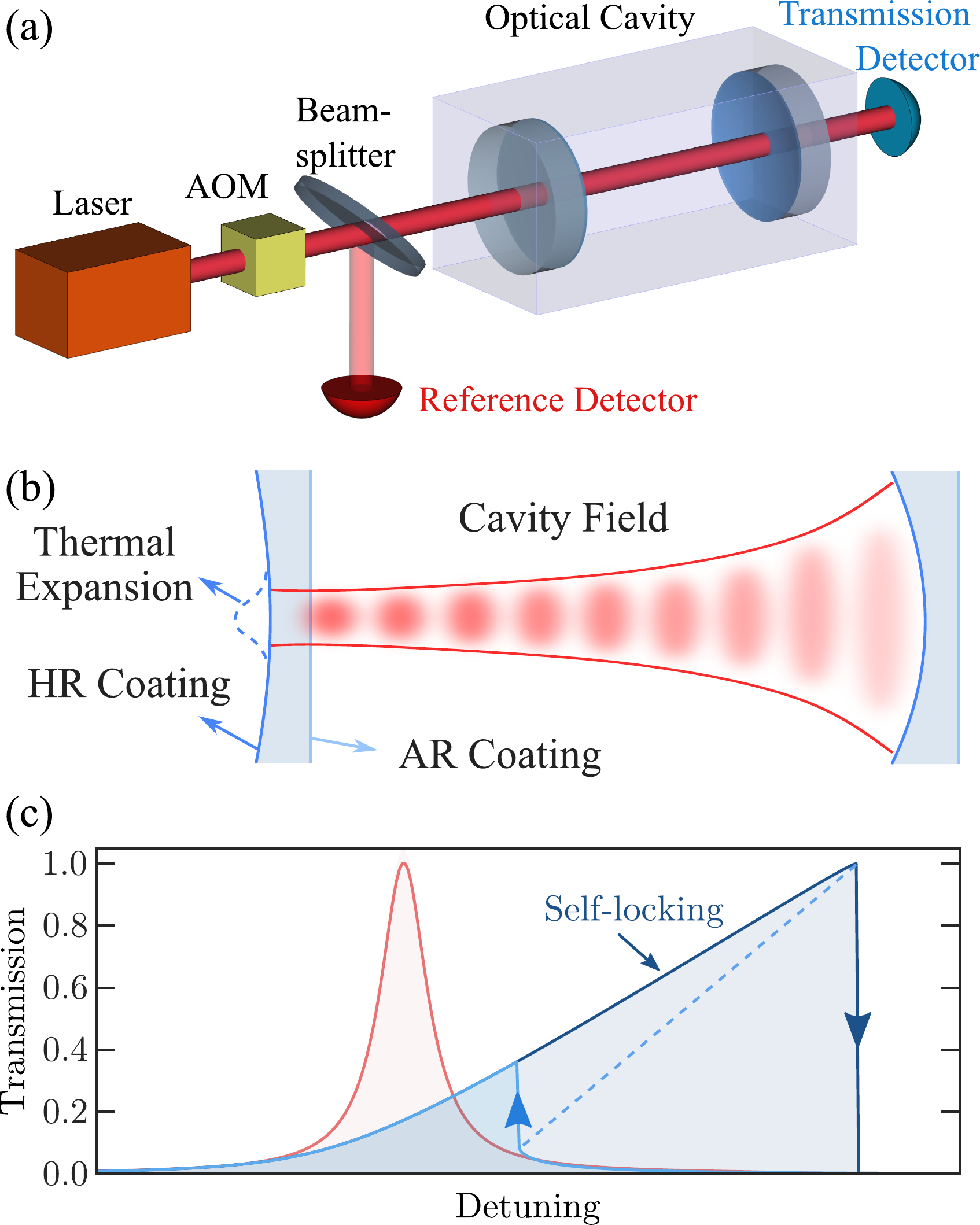} \caption{(a)~Setup. The system enclosed with the blue box is the (generic) cavity to characterize. A low-reflectivity beamsplitter is used to pick up the reference beam before the laser is injected into the system. The input laser power is slightly modulated using an acousto-optic modulator (AOM). The powers of the reference beam and transmitted beam are detected by two photodiodes. (b)~Schematic of the concave-convex optical cavity used for experiment.~The substrate of one of the cavity mirrors is placed inside the cavity to enhance the photothermal effects. (c)~Diagram of system bistability. The red profile shows the typical Lorentzian response of an optical cavity as a function of detuning. In blue, the response of a similar cavity is modified by the photothermal interaction. At high intracavity powers, the system can evolve into the bistable regime where the cavity behavior depends on the scan direction, which drags or skips through resonance depending on the scan direction.}
\label{fig:1_back-action}
\end{figure}

The dominant time-varying signal of the cavity transmission can be normalized as follows by considering the zero-order term of Eq.~(\ref{eq:Qfull}) (see Supplement 1)
 \begin{eqnarray}
t_{\rm norm} & = & \frac{1}{1+\frac{\zeta}{(i\omega/\gamma_{\rm th}-1)}}.  \label{eq:t_norm}
 \end{eqnarray}

\noindent The amplitude and phase of the transmitted signal are then obtained as 
 \begin{eqnarray}
\left| t_{\rm norm} \right|  = \sqrt{\frac{\omega^{2}+\gamma_{\rm th}^{2}}{\omega^{2}+\gamma_{\rm th}^{2}(1-\zeta)^{2}}}, \;
 \phi  =  \arctan \left( \frac{-\omega\zeta\gamma_{\rm th}}{\omega^{2}+(1-\zeta)\gamma_{\rm th}^{2}} \right). \label{eq:t_norm}
 \end{eqnarray}
These quantities can be observed experimentally at different modulation frequencies in order to characterize the photothermal parameters. At low modulation frequencies (i.e., $\omega\ll \gamma_{\rm th}$), the amplitude and phase are predicted to be $\left| t_{\rm norm} \right| = 1\slash{\left|1-\zeta \right|} $ and $\phi = 0$ respectively. At $\omega\gg \gamma_{\rm th}$, the amplitude approximates to one while the phase change relative to the reference modulation tends again to zero. This predicts that one can observe noticeable phase change only for median modulation frequencies. We also note that, replacing the amplitude modulation with a frequency modulation of the driving signal, the resulting formula for the transmitted signal is functionally identical (see Supplement 1). This indicates that one can use either amplitude or phase/frequency modulation to extract the photothermal parameters with the same model. It is worth noting that one can obtain the photothermal relaxation rate $\gamma_{\rm th}$ without any knowledge of other cavity parameters. However, more specific knowledge of the system is required to extract the photothermal coefficient $\beta$ from Eq.~(\ref{eq:zeta}).

\section{Experimental setup}
The proposed scheme for exploring the photothermal effects is shown in Fig.~\ref{fig:1_back-action}(a). The optical cavity, enclosed within a transparent box, is the element to be characterized. The Fabry-P\'erot resonator represented in the figure can be replaced by any other type of optical resonator without loss of generality. The input laser is sent through a modulator before acting as the input to the cavity. We will consider amplitude modulation of the laser beam by an acousto-optic modulator (AOM), even though it is worth reminding that our characterization of the photothermal parameters can also be performed with phase or frequency modulation. A low-reflectivity beamsplitter following the AOM is used to pick up a reference signal. We scan the modulation frequency and vary the effective cavity detuning to collect data from two detectors, one (red detector) as a reference tapped off the driving input, and the other (blue detector) on transmission encoding the photothermal response of the cavity. We obtain the phase response of the cavity by comparing the relative phase difference between the reference and the transmitted signal. 

\begin{figure}[htbp]
\centering
\includegraphics[width=0.47\textwidth]{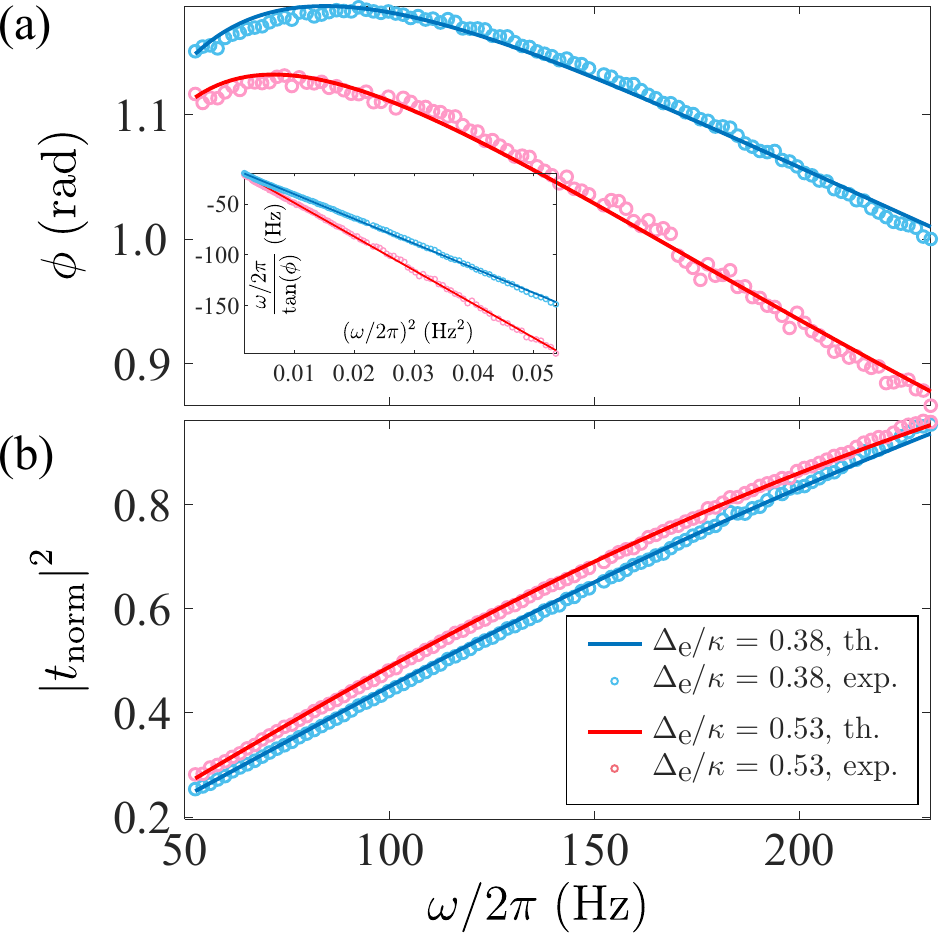} \caption{
(a) Phase response of the cavity as a function of modulation frequency (dots) and corresponding nonlinear fitting (solid lines). A nonlinear transformation of the data is performed to achieve linearity, as shown in the insets. (b) We put the parameters obtained from the fitting of phase into the amplitude response. We still find a good agreement between the theory and experiment.} \label{fig:2_back-action}
\end{figure}

We experimentally observe the optical correction to the photothermal relaxation rate using a concave-convex Fabry P\'erot cavity, as shown in Fig.~\ref{fig:1_back-action}(b) with a finesse of 5700 and a decay rate of \SI{520}{kHz}. The power of the cavity input laser is set as \SI{100}{mW}. The setup is in room-temperature and ambient-pressure conditions, and is built using an Invar spacer to reduce the effect of stochastic thermal fluctuations. Both cavity mirrors are fused-silica substrates with ion-beam sputter coating for high reflectivity at our operating wavelength of \SI{1064}{\nano\meter}. The coating has an optical absorption of less than 10 ppm, offering very small susceptibility to photothermal effects. The substrate of the front mirror is placed inside the cavity so as to enhance the photothermal effects [see Fig.~\ref{fig:1_back-action}(b)]\cite{MaPhotothermally2020}. The mirror coatings expand outwardly, and the refractive index of the substrate changes when the cavity mirrors are heated by the intracavity optical field. Several photothermal effects are present in the system (see Supplement 1 for further discussion). Here we look at their net contributions and assume that different photothermal effects collectively change the cavity response in the same way, either because only one is dominant or because they all cooperate with similar time constants. This approach allows us to measure the effective photothermal parameters that are crucial for the characterization and analysis of cavity photothermal response.

The bistable response occurring in these conditions is represented in Fig.~\ref{fig:1_back-action}(c), which shows the cavity response obtained under strong photothermal interaction as follows from Eqs.~(\ref{eq:steadyq}) and (\ref{eq:steadya_back-action}). The typical Lorentzian response of an optical cavity is deformed, and because of the sign of the interaction the cavity will self-stabilize when in the red-detuning regime~\cite{AnOptical1997, CarmonDynamical2004}. We employ this behavior to explore the system dynamics without the need of any external active feedback control.

We show the measured phase response of the cavity transmission in Fig.~\ref{fig:2_back-action}(a) for two different detunings, $\Delta_{\rm e}/\kappa = 0.38$ (blue) and $\Delta_{\rm e}/\kappa = 0.53$ (red). The data (circles) is fitted according to Eq.~(\ref{eq:t_norm}) using nonlinear regression.~The experimental data is in excellent agreement with the model.~The values of $\gamma_{\rm th}/2\pi$ obtained from the best fit at these two different detunings are rather close, i.e., $16.3\pm 0.2$ \SI{}{\hertz} and $16.0\pm 0.2$ \SI{}{\hertz} respectively. The values of the other free parameter $\zeta$ are $-25.5\pm0.2$ for $\Delta_{\rm e}/\kappa = 0.38$ and $-19.2\pm0.2$ for $\Delta_{\rm e}/\kappa = 0.53$. The errors given here indicate the 95\% confidence interval of the nonlinear regression. We apply a nonlinear transformation (see Supplement 1) to the data to achieve linearity, as shown in the insets of Fig.~\ref{fig:2_back-action} (a). The weighted linear fitting of the transformed data gives compatible results as the nonlinear regression. We substitute the value of $\gamma_{\rm th}$ and $\zeta$ obtained from the best nonlinear fit of phase into Eq.~(\ref{eq:t_norm}) to get the theoretical estimation of the amplitude, indicated by the solid lines of Fig.~\ref{fig:2_back-action}(b). We can also see a good agreement between the data and the model. We note that the amplitude of the transmission is nonlinearly dependent on the modulation frequency, as shown in Eq. (11), even though the dependence looks linear in Fig. 2(b) in the given frequency range. This frequency range is chosen such that the measured phase is more sensitive to frequency change and thus provides smaller uncertainty for the fitting process.

In principle, both the amplitude and the phase signals are suitable candidates for fitting the photothermal response. In practice, however, the amplitude parameter requires a normalization process that involves the calibration of $\varepsilon_l$, $\varepsilon_0$, $\Delta_{\rm e}$, and $\kappa$, or equivalently just an overall normalization factor $N$ (see Supplement 1). Thus, fitting for amplitude adds a layer of complexity that can be readily avoided by considering the phase parameter instead. In Fig.~\ref{fig:2_back-action}, the traces of amplitude obtained at two detunings are rather close while the two plots for the phase are distinctive, confirming that phase fitting is a better candidate to process and extract the photothermal parameters.

\section{Results}
\subsection{Optical correction of the photothermal relaxation rate}
\begin{figure}[htbp]
\centering
\includegraphics[width=0.48\textwidth]{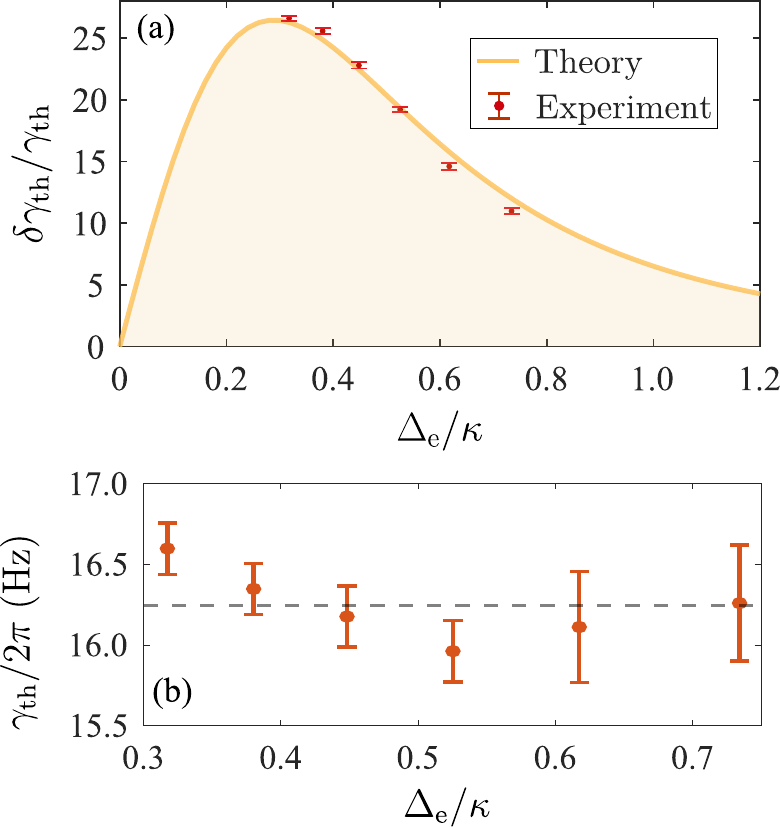} \caption{(a) Optical correction of photothermal relaxation rate as a function of the effective cavity detuning.~The optical correction of the natural photothermal relaxation rate is shown to be nonlinearly dependent on the cavity detuning. The dots with error bars are the experiment results, and the solid line presents the theoretical inference. (b) Photothermal relaxation rate $\gamma_{\rm th}$ characterized at different detunings. The error bars indicate the 95\% confidence bounds of the fitting. The dashed line is the mean value of the measurements. These measurements give us $\gamma_{\rm th}/2\pi= 16.2\pm 0.2$ \SI{}{\hertz}, where the error is the standard deviation of multiple measurements.}
\label{fig:3_back-action}
\end{figure}

Recalling Eq.~(\ref{eq:PTcorrection}), the optical correction effect is manifested in the dimensionless quantity $\zeta$, which can be characterized by fitting the data of cavity transmission into Eq.~(\ref{eq:t_norm}). Fig.~\ref{fig:3_back-action}(a) presents the nonlinear dependence of $\delta\gamma_{\rm th}/\gamma_{\rm th}$ (i.e., $\delta\gamma_{\rm th}/\gamma_{\rm th} = -\zeta$) on the normalized detuning $\Delta_{\rm e}/\kappa$. The results show that the modification of $\gamma_{\rm th}$ in the presence of the cavity field can be tens of times larger than its natural value. This detuning-dependent feature of photothermal response can be crucial in exploring the dynamics of a cavity-based system. For example, the natural photothermal relaxation rate is generally slower than the mechanical response in many optomechanical systems. At specific parameter regimes, however, the photothermal spring can speed up the photothermal effects excitation to the point where the mechanical and photothermal response rates are comparable.

A straightforward application of the optical correction effect is the precise characterization of photothermal parameters. Generally, the photothermal response is relatively slow, and thus its characterization is performed at the low-frequency regime where the data collected is more susceptible to environmental noise and may be limited by the integration time. The cavity-induced optical correction, however, allows us to characterize the photothermal parameters at modulation frequencies much higher than $\gamma_{\rm th}$ to reach high precision. The agreement between data and experiment presented in Fig.~\ref{fig:3_back-action}(a) is an indication of the potential precision of this technique.

\begin{figure*}[htbp]
\centering
\includegraphics[width=0.9\textwidth]{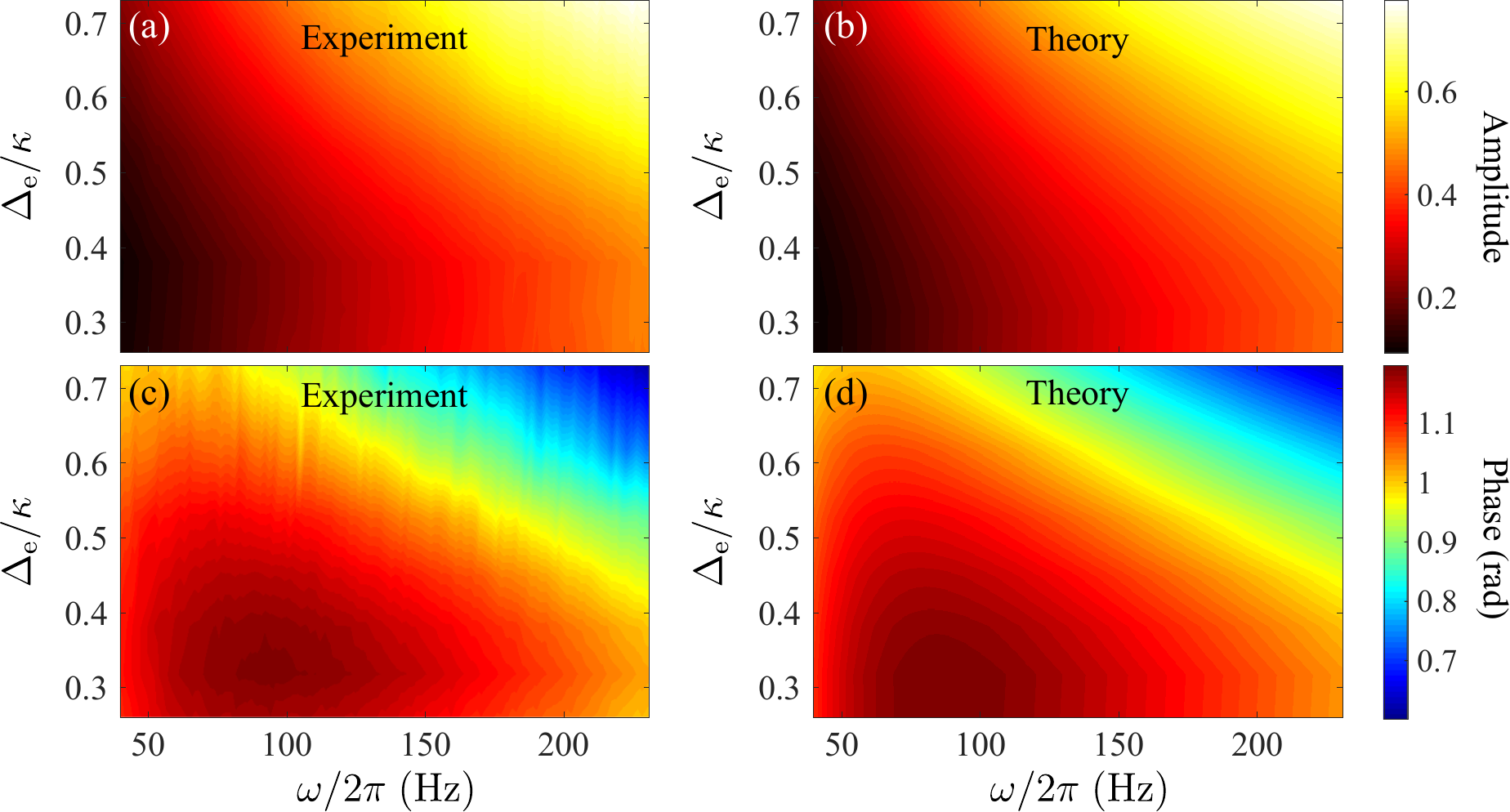} \caption{(a)-(b) The amplitude responses as a function of modulation frequency $\omega$ and effective cavity detuning $\nu = \Delta_{\rm e}/\kappa$. The bandwidth of the photothermal effects increases as the cavity is set close to its resonance. (c)-(d) The phase responses of the cavity.}
\label{fig:5_back-action}
\end{figure*}

\subsection{Precision of characterization}

Our scheme allows us to estimate the value of the photothermal coefficient. As suggested by Eq.~(\ref{eq:zeta}), the parameter $\zeta$ varies only as a function of the normalized detuning $\nu = \Delta_{\rm e}/\kappa$. With the data presented in Fig.~\ref{fig:3_back-action}(a), we employ this feature to do a linear fitting in terms of $\frac{\nu}{(\nu^{2}+1/4)^{2}}$ to obtain the effective cavity photothermal coefficient $\sigma$. The data and its best fit correspond to a coefficient $\sigma=-10.2\pm0.4$. The negative value of $\sigma$, which is proportional to the photothermal coefficient $\beta$, indicates that the effective optical path length of the cavity increases when the optical field heats the cavity mirrors. 

In addition, the fitting values of $\gamma_{\rm th}$ at different detunings are given in Fig.~\ref{fig:3_back-action}(b). These measurements give an average photothermal relaxation rate $\gamma_{\rm th}/2\pi = 16.2\pm 0.2$ \SI{}{\hertz}. The error is the standard deviation over the multiple measurements. To compare the measurement precision of photothermal parameters with previous works~\cite{de_rosa_experimental_2002, BlackEnhancedphotothermaldisplacement2004, farsi_photothermal_2012}, we can infer the effective thermal conductivity using $\kappa_{\rm th} = \gamma_{\rm th}\rho C r_0^2$ \cite{cerdonio_thermoelastic_2001}. Here $r_0$ is the beam radius at the front mirror where the photothermal effects are dominant, and $C$ and $\rho$ are respectively the specific heat capacity and the density of fused silica. Using the room-temperature values of $C = 6.7\times 10^2$ \SI{}{\joule/\kilo\gram.\kelvin}, and $\rho = 2.2\times10^3$ \SI{}{\kilo\gram/\meter^3}, and inferring the beam radius to be $r_0 = 50/\sqrt{2}$ \SI{}{\micro\meter} from the cavity geometry, we obtain the effective thermal conductivity value of $\kappa_{\rm th}=1.182\pm0.016$ \SI{}{\watt/\kelvin.\meter}. Two unique features allow for greater precision: the reduced number of free parameters in our model and the ability to modulate the laser power at frequencies higher than $\gamma_{\rm th}$ thanks to the off-resonance optical correction. 

\subsection{Full-system dynamics}
Figure \ref{fig:5_back-action} displays the full response of the system, comparing experimental data (left panels) to theoretical results (right panels).~With regard to the theoretical plots, we used the values obtained from the fits in Fig.~\ref{fig:2_back-action}. There is a good agreement between the experiment and theory. Figure~\ref{fig:5_back-action}(a)-(b) shows that the cavity acts as a high-pass filter with cut-off frequency tuned by the cavity detuning. In other words, the cavity only allows a high-frequency modulated signal to go through. This is due to the low-pass property of the photothermal effects. At low modulation frequencies, the power fluctuation of the intracavity field can be suppressed by the photothermal back-action. The photothermal effects, however, fail to catch up with the fast response of the cavity field at high-frequency modulations. The natural photothermal relaxation can be modified by the cavity's back-action to induce a higher cut-off frequency. Cavity power and detuning, therefore, coordinate the effective high-pass filter response of the cavity. It is noted that the full high-pass filter is not single-pole as suggested by Eq.~(\ref{eq:t_norm}), and its bandwidth might also be subject to Cerdonio's theory~\cite{cerdonio_thermoelastic_2001}. In Fig.~\ref{fig:5_back-action}(c)-(d), we show that the phase change of the laser going through the cavity due to photothermal effects peaks at a small detuning of about $0.3\kappa$ and a modulation frequency of about 90 \SI{}{\hertz}. 

\section{Conclusion}
We observe that the natural photothermal relaxation rate can be altered in the presence of photothermal-cavity interaction. This feature, analogous to the optical spring effect, can be crucial in analyzing the photothermal dynamics in a cavity-based system and indicates a way of building optical filters with tunable critical cut-off frequency. Also, we report a convenient technique for precisely characterizing photothermal parameters in situ by employing the optical correction effect. Our model shows that one can either modulate the power of the optical input field or the cavity detuning to achieve the characterization of photothermal effects. Experimentally, the measured amplitude and phase of the cavity transmission agree excellently with the theoretical model. The best fit of the phase response gives the photothermal relation rate of $16.2\pm0.2$ \SI{}{\hertz}. This characterization is an order of magnitude more precise than previous works. It is worth noting that the natural photothermal relaxation rate of a cavity mirror depends on the laser power density, and the photothermal coefficient is mainly determined by the absorption in the mirror coatings. Appropriate engineering of the mirror coating, substrate material or cavity geometry, therefore, can allow us to observe photothermal optical correction at relatively low laser powers.

\section*{Funding Information}
This research was funded by the Australian Research Council Centre of Excellence CE110001027 and the Australian Government Research Training Program Scholarship. PKL acknowledges support from the ARC Laureate Fellowship FL150100019.

%
%

\section*{Disclosures}
%
%
%
%

\medskip

The authors declare no conflicts of interest.

\section*{Supplemental Documents}
See Supplement 1 for supporting content.




\clearpage

\section*{Supplemental Information}
This document provides supplementary information to ``Optical back-action on the photothermal relaxation rate''. Section 1 details the calculations for the equation of motions in the case of amplitude modulation. Section 2 provides a more accurate analytical solution with the dimensionless approach. In section 3, we show that the formula of normalized cavity transmission is equivalent in the case of either the amplitude or phase modulation of the cavity field. Section 4 discusses the primary contribution of different photothermal effects in our experimental system.

\subsection*{1. Modulation of Amplitude}
Photothermal displacement $q_{\rm th}$ can modulate the cavity resonance eigenfrequency $\omega_{\rm c}$ via changing the cavity length $L_{\rm c}$: 
\begin{eqnarray*}
\omega_{\rm c}(q_{\rm th}) & = & \frac{\text{n\ensuremath{\pi c}}}{L_{\rm c}+q_{\rm th}}\\
 & \approx & \frac{\text{n\ensuremath{\pi c}}}{L_{\rm c}}+\frac{\text{n\ensuremath{\pi c}}}{L_{\rm c}^{2}}q_{\rm th}+...\\
 & = & \omega_{\rm c}+\frac{\omega_{c}}{L_{c}}q_{\rm th}+...\\
 & = & \omega_{\rm c}+Gq_{\rm th}+...
\end{eqnarray*}
Here we assume $q_{\rm th} \ll L_{\rm c}$. We therefore define the parameter $G = \omega_{c}/L_{c}$ as the coupling rate of photothermal-cavity interaction. Denoting by $a$ the amplitude of the intracavity field, we have that the dynamics of the photothermal interaction between the two can be modelled by the following two equations of motion in the rotating frame of frequency $\omega_l$ of the input laser,
\begin{eqnarray}
 \dot{q}_\textrm{\rm th} & = &-\gamma_{\rm th}(q_\textrm{\rm th}+\beta P_{c}) \label{eq:qth} ,\\
 \dot{a}& = & -[\kappa/2-i(\Delta+Gq_\textrm{\rm th})]a+\varepsilon_{\rm l}+\varepsilon_{\rm 0}\cos(\omega t+\text{\ensuremath{\varphi}}), \label{eq:a}
\end{eqnarray}
where $\gamma_{\rm th}$ is the photothermal relaxation rate, $\beta$ is the photothermal response coefficient and $\kappa$ denotes the total loss of the cavity. The term $iGq_\textrm{\rm th}a$ in Eq.~(\ref{eq:a}) indicates the coupling between cavity mode and photothermal effects at the rate $G$. The intracavity power is written explicitly as $P_{c}=\hbar\omega_{\rm c}\left\vert a\right\vert ^{2}/\tau_{\rm cav} = \hbar G c\left\vert a\right\vert ^{2}/2$ with $\tau_{\rm cav}=2L_{c}/c$ being the round-trip time of cavity photons. The cavity is driven at a detuning $\Delta$ from resonance, by an optical field of amplitude $\epsilon_\textrm{l}$ sinusoidally modulated at a frequency $\omega$ and phase $\varphi$ by a percentage $\varepsilon_0/\varepsilon_\textrm{l}\ll1$. Note that the modulation is external and applied on purpose for characterizing the photothermal effects.

As the modulation is small, we can assume small deviations from the steady-state solutions and substitute the assumptions $q_\textrm{\rm th}=q_\textrm{\rm th}^{0}+\delta q_\textrm{\rm th}$ and $a=a_{0}+\delta a$ into Eqs.~(\ref{eq:qth})-(\ref{eq:a}) to obtain the steady states of the system:
\begin{eqnarray}
0 & = & q_\textrm{\rm th}^{0}+\alpha\left\vert a_{0}\right\vert ^{2} \label{eq:steadyq}\\
0 & = & -\kappa a_{0}/2+i(\Delta a_{0}+Ga_{0}q_\textrm{\rm th}^{0})+\varepsilon_{\rm l}, \label{eq:steadya}
\end{eqnarray}
and the first-order dynamics:
\begin{eqnarray}
\delta \dot{q}_\textrm{\rm th} & = & -\gamma_{\rm th}[\delta q_\textrm{\rm th}+\alpha(a_{0}\delta a^{*}+a_{0}^{*}\delta a)] \label{eq:dq}\\
\delta \dot{a} & = & -\kappa\delta a/2+i\Delta\delta a+iG(a_{0}\delta q_\textrm{\rm th}+q_\textrm{\rm th}^{0}\delta a).\nonumber\\
&  & +\frac{\varepsilon_{\rm 0}}{2}e^{-i(\omega t+\varphi)}+\frac{\varepsilon_{\rm 0}}{2}e^{i(\omega t+\varphi)}, \label{eq:da}
\end{eqnarray}
where $\alpha=\hbar G c\beta/2$.

To solve the first-order dynamical equations [i.e., Eqs.~(\ref{eq:dq})-(\ref{eq:da})], we use the following ansatz:
\begin{eqnarray}
\delta q_\textrm{\rm th} & = & Q e^{-i\omega t}+Q^{*}e^{i\omega t} \nonumber\\
\delta a & = & A_{-}e^{-i\omega t}+A_{+}e^{i\omega t}\\
\delta a^{*} & = & A_{-}^{*}e^{i\omega t}+A_{+}^{*}e^{-i\omega t} \nonumber
\end{eqnarray}
We obtain the following equations for the amplitudes of the first-order sidebands.
 \begin{eqnarray}
(i\omega-\gamma_{\rm th})Q & = & \gamma_{\rm th}\alpha[a_{0}A_{+}^{*}+a_{0}^{*}(A_{-})] \label{eq:qw}\\
\text{[}\kappa/2-i(\Delta_{e}+\omega)]A_{-} & = & iGa_{0}Q+\frac{\varepsilon_{\rm 0}}{2}e^{-i\varphi} \label{eq:ANw}\\ \text{[}\kappa/2+i(\Delta_{e}-\omega)]A_{+}^{*} & = & -iGa_{0}^{*}Q+\frac{\varepsilon_{\rm 0}}{2}e^{-i\varphi} \label{eq:APw}
\end{eqnarray}

where $\Delta_e = \Delta+Gq_\textrm{\rm th}^0$ is the effective cavity detuning. Since the modulation frequency is determined specifically for the modulation, we can set it so that $\omega\ll\Delta_e$. Under this condition, we substitute Eqs.~(\ref{eq:ANw})-(\ref{eq:APw}) into Eq.~(\ref{eq:qw}) and obtain the solution for the photothermal displacement:

\begin{eqnarray*}
(i\omega-\gamma_{\rm th})Q & = & \gamma_{\rm th}\alpha[\frac{a_{0}(-iGa_{0}^{*}Q+\frac{\varepsilon_{0}}{2}e^{-i\varphi})}{i\Delta_{e}+\kappa/2}+\frac{a_{0}^{*}(iGa_{0}Q+\frac{\varepsilon_{0}}{2}e^{-i\varphi})}{-i\Delta_{e}+\kappa/2}]\\
 & = & \gamma_{\rm th}\alpha\frac{a_{0}(-iGa_{0}^{*}Q+\frac{\varepsilon_{0}}{2}e^{-i\varphi})(-i\Delta_{e}+\kappa/2)+a_{0}^{*}(iGa_{0}Q+\frac{\varepsilon_{0}}{2}e^{-i\varphi})(i\Delta_{e}+\kappa/2)}{\Delta_{e}^{2}+\kappa^{2}/4}\\
 & = & \gamma_{\rm th}\alpha\frac{i\Delta_{e}\frac{\varepsilon_{0}}{2}e^{-i\varphi}(a_{0}^{*}-a_{0})+\frac{\varepsilon_{0}}{2}e^{-i\varphi}\kappa/2(a_{0}^{*}+a_{0})-2G\left|a_{0}\right|^{2}\Delta_{e}Q}{\Delta_{e}^{2}+\kappa^{2}/4}\\
\end{eqnarray*}
\begin{eqnarray}
[(i\omega-\gamma_{\rm th})(\Delta_{e}^{2}+\kappa^{2}/4)+2\varepsilon\alpha G\left|a_{0}\right|^{2}\Delta_{e}]Q & = & \gamma_{\rm th}\alpha\frac{\varepsilon_{0}}{2}e^{-i\varphi}[i\Delta_{e}(a_{0}^{*}-a_{0})+\kappa/2(a_{0}^{*}+a_{0})] \nonumber\\
 & = & \gamma_{\rm th}\alpha\varepsilon_{l}\frac{\varepsilon_{0}}{2}e^{-i\varphi}(i\Delta_{e}\frac{-2i\Delta_{e}}{\Delta_{e}^{2}+\kappa^{2}/4}+\kappa/2\frac{\kappa}{\Delta_{e}^{2}+\kappa^{2}/4}) \nonumber\\
 & = & \gamma_{\rm th}\alpha\varepsilon_{l}\varepsilon_{0}e^{-i\varphi} \nonumber
\end{eqnarray}

\begin{eqnarray}
Q & = & \frac{\gamma_{\rm th}\alpha\varepsilon_{l}\varepsilon_{0}e^{-i\varphi}}{(i\omega-\gamma_{\rm th})(\Delta_{e}^{2}+\kappa^{2}/4)+2\varepsilon\alpha G\left|a_{0}\right|^{2}\Delta_{e}} \nonumber\\
 & = &\frac{\gamma_{\rm th}\alpha\varepsilon_{\rm l}\varepsilon_{\rm 0}e^{-i\varphi}}{(\Delta_{\rm e}^{2}+\kappa^{2}/4)}\cdot\frac{1}{i\omega-\gamma_{\rm th}+\zeta\gamma_{\rm th}} \label{eq:qsol}
\end{eqnarray}
with
  \begin{eqnarray}
\zeta &=&\frac{2\alpha G\left|a_{0}\right|^{2}\Delta_{e}}{\Delta_{e}^{2}+\kappa^{2}/4} \label{eq:zeta}
 \end{eqnarray}

Considering the input-output relation of an impedance-matched optical cavity, the cavity transmission is
\begin{eqnarray}
\left|t\right|^{2} & = & \left| \frac{\kappa a}{2}\right|^{2} \nonumber\\
 & = & \frac{\kappa^{2}}{4}[a_{0}a_{0}^{*}+A_{-}A_{-}^{*}+A_{+}A_{+}^{*} \nonumber\\
 &  & +(a_{0}^{*}A_{-}+a_{0}A_{+}^{*})e^{-i\omega_{p}t}+(a_{0}A_{-}^{*}+a_{0}^{*}A_{+})e^{i\omega_{p}t} \nonumber\\
 &  & +A_{-}A_{+}^{*}e^{-2i\omega_{p}t}+A_{+}A_{-}^{*}e^{2i\omega_{p}t}]
\end{eqnarray}

Since the amplitude is much smaller for second-order sidebands than first-order ones, the dominant time-varying signal of the cavity transmission is expressed as follows by considering Eqs.~(\ref{eq:qw}) and (\ref{eq:qsol})
\begin{eqnarray}
t_{1} & = & \frac{\kappa^2}{4}[2(a_{0}A_{+}^{*}+a_{0}^{*}A_{-})] \nonumber\\
 & = & \frac{\kappa^2}{4}\frac{2(i\omega-\gamma_{\rm th})}{\gamma_{\rm th}\alpha}Q \nonumber\\
 & = & \frac{\kappa^2}{4}\frac{2\varepsilon_{\rm l}\varepsilon_{\rm 0}e^{-i\varphi}{/(\Delta_{e}^{2}+\kappa^{2}/4)}}{1+\frac{\zeta\gamma_{\rm th}}{i\omega-\gamma_{\rm th}}} \label{eq:t1}
\end{eqnarray}
The equation above can be normalized with the normalization factor of $2\varepsilon_{\rm l}\varepsilon_{\rm 0}{/(\Delta_{e}^{2}+\kappa^{2}/4)}$ as
 \begin{eqnarray}
t_{\rm norm} & = & \frac{1}{1+\frac{\zeta}{(i\omega/\gamma_{\rm th}-1)}}  \label{eq:t_norm}
 \end{eqnarray}
The amplitude and phase of the transmitted signal are then easily obtained as
\begin{eqnarray}
\left| t_{\rm norm} \right| & = & \sqrt{\frac{\omega^{2}+\gamma_{\rm th}^{2}}{\omega^{2}+\gamma_{\rm th}^{2}(1-\zeta)^{2}}}  \label{eq:amp}\\
 \tan\phi & = & \frac{-\omega\zeta\gamma_{\rm th}}{\omega^{2}+\gamma_{\rm th}^{2}-\zeta\gamma_{\rm th}^{2}} \label{eq:phase}
\end{eqnarray}

We can also express Eqs. (\ref{eq:amp})-(\ref{eq:phase}) as linear equations,
\begin{eqnarray}
y_a = a_1 x_a + a_2 \label{eq:la}\\
y_p = p_1 x_p + p_2 \label{eq:lp}
\end{eqnarray} 
where
\begin{eqnarray}
x_a &=& \omega^{2};  \;\;\;\;\;\;\;\;\;\;\;\;\;\;\;\; 
y_a = \frac{1}{|t_{\rm norm}|^{2}-1} \nonumber\\
a_1 &=& \frac{1}{(2\zeta-\zeta^{2})\gamma_{\rm th}^{2}}; \;\;\;
a_2 = \frac{(1-\zeta)^{2}}{2\zeta-\zeta^{2}}
\end{eqnarray} 
and
\begin{eqnarray}
x_p &=&  \omega^{2}; \; \; \; 
y_p = \frac{\omega}{\tan\phi} \nonumber \\
p_1 &=&  \frac{-1}{\zeta\gamma_{\rm th}}; \;  \; 
p_2 = \frac{\zeta\gamma_{\rm th}-\gamma_{\rm th}}{\zeta} \label{eq:lpt}
\end{eqnarray} 

We can rewrite Eq.~(\ref{eq:zeta}) as follows,
\begin{eqnarray}
\zeta &=& \frac{\sigma\nu}{(\nu^{2}+1/4)^{2}} \label{eq:newzeta}
\end{eqnarray} 
where $\sigma = 2\alpha\varepsilon_{\rm l}^{2}G/\kappa^{3}$ and $\nu = \Delta_e/\kappa$. It is reasonable to define the dimensionless physical quantity $\sigma$ as an effective photothermal coefficient. If plotting $\zeta$ as a function of $\frac{\nu}{(\nu^{2}+1/4)^{2}}$ , the slope of the linear fitting gives the value of $\sigma$. 

The normalized Lorentzian profile of a cavity response is given by
\begin{eqnarray}
C = \frac{1/4}{(\nu^2+1/4)}
\end{eqnarray} 
Its first derivative is easily obtained as
\begin{eqnarray}
C^\prime &=& \frac{\nu/2}{(\nu^{2}+1/4)^{2}} 
\end{eqnarray} 
which is linear to the photothermal modification term $\zeta$ defined in \eqref{eq:zeta} and \eqref{eq:newzeta}. This result suggests that the optical modification of the photothermal time constant is proportional to the slope of the cavity Lorentzian response. The maximum slope and the maximum modification are both found at $\nu = 1/2\sqrt{3} \simeq 0.29$.

\subsection*{2. A more accurate analytical approximation with dimensionless approach}
We consider a cavity with photothermal expansion, which has equations of motion
\begin{equation}
  \begin{aligned}
    \dot{q}_{\rm th}       &= -\gamma_{\rm th}\left[q_{\rm th}+\beta\frac{\hbar\omega_c}{t_c}|a|^2\right], \\
    \dot{a}  &= i(\Delta+Gq_{\rm th})a-\frac{\delta\omega}{2}a+\varepsilon_l+\varepsilon_0\cos(\omega t+\phi).
  \end{aligned}
\end{equation}
The number of parameters can be reduced through nondimensionalisation. We introduce a natural length scale $\ell$ corresponding to the distance of photothermal expansion which will shift the cavity resonance by one linewidth
\begin{equation}
  \ell=\frac{\delta\omega/2}{G},
\end{equation}
and take as natural frequency the photothermal relaxation rate $\gamma_{\rm th}$. Our dimensionless time will then be $\tilde{\tau}=\gamma_{\rm th} t$. We also introduce the maximum cavity amplitude due to the main laser $\varepsilon_l$:
\begin{equation}
  \tilde{A}=\frac{\varepsilon_l}{\delta\omega/2}.
\end{equation}
The dimensionless dynamical variables are then
\begin{equation}
  \tilde{q}_{\rm th}=\frac{q_{\rm th}}{\ell},\tilde{a}=\frac{a}{\tilde{A}}.
\end{equation}
In terms of these the equations of motion become
\begin{equation}\label{eq:om:PTOnlyNDEOM}
  \begin{aligned}
    \tilde{q}_{\rm th}'      &= -(\tilde{q}_{\rm th}+\tilde{\beta}|\tilde{a}|^2), \\
    \tilde{a}' &= \tilde{\eta}\left(i(\tilde{\Delta}+\tilde{q}_{\rm th})\tilde{a}-\tilde{a}+1+\tilde{\varepsilon}_0\cos(\tilde{\omega}\tilde{\tau}+\phi)\right).
  \end{aligned}
\end{equation}
In Eq.~(\ref{eq:om:PTOnlyNDEOM}) primes represent derivatives with respect to $\tilde{\tau}$, and we have we introduced the dimensionless parameters
\begin{equation}
  \tilde{\beta}=\frac{\beta\hbar\omega\tilde{A}^2}{\ell t_c}=\frac{\beta\hbar\omega\varepsilon_l^2}{t_c}\frac{G}{(\delta\omega/2)^3};\;\tilde{\eta}=\frac{\delta\omega/2}{\gamma_{\rm th}};\;\tilde{\varepsilon}_0=\frac{\varepsilon_0}{\tilde{A}(\delta\omega/2)}.
\end{equation}
The quantity $\tilde{\omega}=\omega/\gamma_{\rm th}$ represents the frequency as a multiple of the photothermal relaxation rate, and $\tilde{\Delta}=\Delta/(\delta\omega/2)$ gives the detuning measured in cavity linewidths. We bring particular attention to $\tilde{\eta}$, the ratio of the optical timescale to the photothermal relaxation rate. This will typically be very large, of order greater than $10^3$ while all other dimensionless parameters have been scaled to be of order $1$, and we will later exploit this and expand to first order around $1/\tilde{\eta}\approx 0$.

We next linearise the dimensionless equations of motion by assuming $\tilde{q}_{\rm th}=\tilde{q}_{\rm th,s}+\delta\tilde{q}_{\rm th}$ and $\tilde{a}=\tilde{a}_s+\delta\tilde{a}$, where $\tilde{q}_{\rm th,s}$, $\tilde{a_s}$ are the steady states of Eq.~(\ref{eq:om:PTOnlyNDEOM}) when $\tilde{\varepsilon}_0=0$. Then to first order
\begin{equation}
  \begin{aligned}
    \delta\tilde{q}_{\rm th}' &= -\left[\tilde{q}_{\rm th,s}+\delta\tilde{q}_{\rm th}+\tilde{\beta}(\tilde{a}_s^*+\delta\tilde{a}^*)(\tilde{a}_s+\delta\tilde{a})\right], \\
                     &= -\left[\left(\tilde{q}_{\rm th,s}+\tilde{\beta}|\tilde{a}_s|^2\right)+\delta\tilde{q}_{\rm th}+\tilde{\beta}(\tilde{a}_s^*+\delta\tilde{a}^*)(\tilde{a}_s+\delta\tilde{a})\right], \\
      &=-\left[\delta\tilde{q}_{\rm th}+\tilde{\beta}(\tilde{a}_s^*\delta\tilde{a}+\tilde{a}_s\delta\tilde{a}^*)\right].
  \end{aligned}
\end{equation}
Proceeding similarly for $\delta\tilde{a}'$ and expanding the cosine as a sum of exponentials, we find the linear equations of motion
\begin{equation}\label{eq:om:PTOnlyLin}
  \begin{aligned}
    \delta\tilde{q}_{\rm th}'      &= -\left[\delta\tilde{q}_{\rm th}+\tilde{\beta}(\tilde{a}_s^*\delta\tilde{a}+\tilde{a}_s\delta\tilde{a}^*)\right], \\
    \delta\tilde{a}' &= \tilde{\eta}\left[i(\tilde{\Delta}+\tilde{q}_{\rm th,s})\delta\tilde{a}+i\tilde{a}_s\delta\tilde{q}_{\rm th}-\delta\tilde{a}+\frac{\tilde{\varepsilon}_0}{2}\left(e^{i(\tilde{\omega}\tilde{\tau}+\phi)}+e^{-i(\tilde{\omega}\tilde{\tau}+\phi)}\right)\right].
  \end{aligned}
\end{equation}

To solve Eq.~(\ref{eq:om:PTOnlyLin}) we introduce the ansatz
\begin{equation}
  \begin{aligned}
    \delta\tilde{q}_{\rm th} &= \tilde{Q}e^{-i\tilde{\omega}\tilde{\tau}}+\tilde{Q}^*e^{i\tilde{\omega}\tilde{\tau}}, \\
    \delta\tilde{a}    &= \tilde{A}_-e^{-i\tilde{\omega}\tilde{\tau}}+\tilde{A}_+e^{i\tilde{\omega}\tilde{\tau}}, \\
    \delta\tilde{a}^*  &= \tilde{A}_-^*e^{i\tilde{\omega}\tilde{\tau}}+\tilde{A}_+^*e^{-i\tilde{\omega}\tilde{\tau}},
  \end{aligned}
\end{equation}
where the coefficients $\tilde{Q}$, $\tilde{A}_{\pm}$ are unknown quantities to be solved for. Substituting these into the linearised equation for $\delta\tilde{q}_{\rm th}'$, we find
\begin{equation}
  \begin{aligned}
  &-i\tilde{\omega}\tilde{Q}e^{-i\tilde{\omega}\tilde{\tau}}+i\tilde{\omega}\tilde{Q}^*e^{i\tilde{\omega}\tilde{\tau}} \\
    &\phantom{=}=-e^{-i\tilde{\omega}\tilde{\tau}}\left[\tilde{Q}+\tilde{\beta}\left(\tilde{a}_s^*\tilde{A}_-+\tilde{a}_s\tilde{A}_+^*\right)\right]-e^{i\tilde{\omega}\tilde{\tau}}\left[\tilde{Q}^*+\tilde{\beta}\left(\tilde{a}_s^*\tilde{A}_++\tilde{a}_s\tilde{A}_-^*\right)\right].
  \end{aligned}
\end{equation}
Equating terms oscillating at $e^{i\tilde{\omega}\tilde{\tau}}$ then gives
\begin{equation}\label{eq:om:PTZetaEqn}
  (i\tilde{\omega}-1)\tilde{Q}=\tilde{\beta}(\tilde{a}_s^*\tilde{A}_-+\tilde{a}_s\tilde{A}_+),
\end{equation}
with the $e^{-i\tilde{\omega}\tilde{\tau}}$ terms giving the same equation.
Proceeding similarly for $\delta\tilde{a}'$ yields two relations
\begin{equation}\label{eq:om:PTAPMEqn}
  \begin{aligned}
    \tilde{A}_-\left(i\tilde{\omega}+\tilde{\eta}\left[i\tilde{\Delta}_{\rm e}-1\right]\right)&=-\tilde{\eta}\left(i\tilde{a}_s\tilde{Q}+\frac{\tilde{\varepsilon}_0}{2}e^{-i\phi}\right), \\
    \tilde{A}_+\left(-i\tilde{\omega}+\tilde{\eta}\left[i\tilde{\Delta}_{\rm e}-1\right]\right)&=-\tilde{\eta}\left(i\tilde{a}_s\tilde{Q}^*+\frac{\tilde{\varepsilon}_0}{2}e^{i\phi}\right),
  \end{aligned}
\end{equation}
where we have defined
\begin{equation}
  \tilde{\Delta}_{\rm e}=\tilde{\Delta}-\tilde{q}_{\rm th,s}.
\end{equation}
We can use Eqs.~(\ref{eq:om:PTZetaEqn}) and (\ref{eq:om:PTAPMEqn}) to solve for $\tilde{Q}$, however this expression is somewhat complicated:

\begin{equation}
  \tilde{Q}=\frac{\tilde{\beta}e^{-i\phi}\tilde{\varepsilon}_0\tilde{\eta}\left(\tilde{\eta}(1+\tilde{\Delta}_{\rm e}^2)-i\tilde{\omega}\right)}{(1+\tilde{\Delta}_{\rm e}^2)\left((1-i\tilde{\omega})\tilde{\omega}^2+2\tilde{\eta}\tilde{\omega}(i+\tilde{\omega})\right)+\tilde{\eta}^2\left(2\tilde{\beta}\tilde{\Delta}_{\rm e}+i(1+\tilde{\Delta}_{\rm e}^2)^2(i+\tilde{\omega})\right)}.
\end{equation}

To simplify this we recall that the dimensionless parameter $\tilde{\eta}$ will be many orders of magnitude larger than the others, so we may expand in powers of $1/\tilde{\eta}$:
\begin{equation}\label{eq:om:ZetaFirstOrder}
  \tilde{Q}=\frac{\tilde{\beta}e^{-i\phi}\tilde{\varepsilon}_0}{1+\tilde{\Delta}_{\rm e}^2}\frac{1}{i\tilde{\omega}-1+\tilde{\zeta}}\left(1+\frac{1}{\tilde{\eta}}\frac{i\tilde{\omega}}{1+\tilde{\Delta}_{\rm e}^2}\frac{i\tilde{\omega}-1-\tilde{\zeta}}{i\tilde{\omega}-1+\tilde{\zeta}}\right)+\mathcal{O}\left(\frac{1}{\tilde{\eta}^2}\right),
\end{equation}
where we have defined
\begin{equation}
  \tilde{\zeta}=\frac{2\tilde{\beta}\tilde{\Delta}_{\rm e}}{(1+\tilde{\Delta}_{\rm e}^2)^2}=\frac{2\beta G\hbar\omega_c\varepsilon_l^2\Delta_{\rm e}}{t_c\left((\delta\omega/2)^2+\Delta_{\rm e}^2\right)^2}.
\end{equation}
This can be re-dimensionalised as
\begin{eqnarray}
  \begin{aligned}
    Q &= \ell\tilde {Q}, \\
      &= \frac{\gamma_{\rm th}\alpha\varepsilon_{\rm l}\varepsilon_{\rm 0}e^{-i\varphi}}{(\Delta_{\rm e}^{2}+\kappa^{2}/4)}\frac{1}{i\omega-\gamma_{\rm th}+\zeta\gamma_{\rm th}}\times\cdots \\
      &\phantom{=}\left(1+\frac{\gamma_{\rm th}}{\kappa/2}\frac{i\omega}{\gamma_{\rm th}(\Delta_{\rm e}^2+\kappa^2/4)}\frac{i\omega-\gamma_{\rm th}-\gamma_{\rm th}\zeta}{i\omega-\gamma_{\rm th}+\gamma_{\rm th}\zeta}\right)+\mathcal{O}\left(\frac{\gamma_{\rm th}}{\kappa/2}\right)^2,
  \end{aligned}
\end{eqnarray}
and we remark that $\tilde{\zeta}$ is equal to the $\zeta$ defined in the main text.

Looking at the zeroth order term:
\begin{equation}
  \tilde{Q}=\frac{\tilde{\beta}e^{-i\phi}\tilde{\varepsilon}_0}{1+\tilde{\Delta}_{\rm e}^2}\frac{1}{i\tilde{\omega}-1+\tilde{\zeta}}+\mathcal{O}\left(\frac{1}{\tilde{\eta}}\right),
\end{equation}
we see that this represents the cutoff frequency $\gamma_{\rm th}$ (which is $1$ in the dimensionless units) being modified by $\tilde{\zeta}$.

We can use Eq.~(\ref{eq:om:ZetaFirstOrder}) to calculate the cavity transmission. This is given by
\begin{equation}
  \begin{aligned}
    |t|^2 &= \left\lvert\frac{\delta\omega}{2}a\right\rvert^2, \\
      &= \left|\frac{\delta\omega}{2}\tilde{A}\right|^2|\tilde{t}|^2,
  \end{aligned}
\end{equation}
in terms of the dimensionless transmission 
\begin{equation}
  \tilde{t}=\tilde{a}.
\end{equation}
Expanding this:
\begin{equation}
  \begin{aligned}
    |\tilde{t}|^2 &= \left(|\tilde{a}_s|^2+|\tilde{A}_-|^2+|\tilde{A}_+|^2\right. \\
      &\phantom{=}+e^{-i\tilde{\omega}\tilde{\tau}}\left(\tilde{a}_s^*\tilde{A}_-+\tilde{a}_s\tilde{A}_+^*\right)+e^{i\tilde{\omega}\tilde{\tau}}\left(\tilde{a}_s\tilde{A}_-^*+\tilde{a}_s^*\tilde{A}_+\right) \\
      &\phantom{=}\left.+e^{-2i\tilde{\omega}\tilde{\tau}}A_-A_+^*+e^{2i\tilde{\omega}\tilde{\tau}}A_+A_-^*\right).
  \end{aligned}
\end{equation}
The second order sidebands oscillating at $\pm2\tilde{\omega}$ will have amplitude much smaller than the first-order due to the $\tilde{a}_s$ factor and small size of the perturbations $\delta\tilde{a}$, so we can neglect them. The expression for the positive sideband transmission is then
\begin{equation}
  \tilde{t}_1=\tilde{a}_s\tilde{A}_-^*+\tilde{a}_0^*\tilde{A}_+.
\end{equation}
Using Eq.~(\ref{eq:om:PTZetaEqn}) this simplifies to
\begin{equation}
  \begin{aligned}
    \tilde{t}_1 &= \frac{(i\tilde{\omega}-1)\tilde{Q}}{\tilde{\beta}}, \\
      &= \frac{e^{-i\phi}\tilde{\varepsilon}_0}{1+\tilde{\Delta}_{\rm e}^2}\cdot\frac{1}{1+\frac{\tilde\zeta}{i\tilde{\omega}-1}}\left[1+\frac{1}{\tilde{\eta}}\frac{i\tilde{\omega}}{1+\tilde{\Delta}_{\rm e}^2}\frac{1-\frac{\tilde{\zeta}}{i\tilde{\omega}-1}}{1+\frac{\tilde{\zeta}}{i\tilde{\omega}-1}}\right]+\mathcal{O}\left(\frac{1}{\tilde{\eta}^2}\right).
  \end{aligned}
\end{equation}

\subsection*{3. Modulation of Phase}
To characterize the photothermal parameters of an optical cavity including photothermal effects, we can modulate the frequency of the input laser or the cavity length instead of the laser power. The equations of motion can be written as follows:
\begin{eqnarray}
\dot{q_\textrm{\rm th}}  & = & -\gamma_{\rm th}(q_\textrm{\rm th}+\beta P_{c})\\
\dot{a}  &=& -[\kappa/2-i(\Delta+\Delta_{1}\cos(\omega t+\text{\ensuremath{\varphi}})+Gq_\textrm{\rm th})]a+\varepsilon_{\rm l}.
\end{eqnarray}
where $\Delta_1$, $\omega$ and $\varphi$ are the amplitude, frequency and phase of the frequency modulation. The definition of other symbols are the same as the those in Eq.~(\ref{eq:qth})-(\ref{eq:a}).

If taking the assumptions $q_\textrm{\rm th}=q_\textrm{\rm th}^{0}+\delta q_\textrm{\rm th}$,$a=a_{0}+\delta a$, we then have the steady states as follows:
\begin{eqnarray}
0 & = & q_\textrm{\rm th}^{0}+\alpha\left\vert a_{0}\right\vert ^{2}\\
0 & = & -\kappa a_{0}/2+i(\Delta a_{0}+Ga_{0}q_\textrm{\rm th}^{0})+\varepsilon_{\rm l}.
\end{eqnarray}
where $\alpha=\beta\hbar\omega_{c}/\tau_{c}$

The dynamical equations of the first-order are:

\begin{eqnarray}
\dot{\delta q_\textrm{\rm th}} & = & -\gamma_{\rm th}[\delta q_\textrm{\rm th}+\beta\hbar\omega_{c}(a_{0}\delta a^{*}+a_{0}^{*}\delta a)/\tau_{c}]\\
\dot{\delta a} & = & -\kappa\delta a/2+i\Delta_{0}\delta a+iG(a_{0}\delta q_\textrm{\rm th}+q_\textrm{\rm th}^{0}\delta a) \nonumber\\
&  & +i(\frac{\Delta_{1}a_{0}}{2}e^{-i(\omega t+\varphi)}+\frac{\Delta_{1}a_{0}}{2}e^{i(\omega t+\varphi)}).
\end{eqnarray}

We substitude the ansatz
$\delta q_\textrm{\rm th}  =  Q e^{-i\omega t}+Q ^{*}e^{i\omega t}$, 
$\delta a  =  A_{-}e^{-i\omega t}+A_{+}e^{i\omega t}$, and
$\delta a^{*}  =  A_{-}^{*}e^{i\omega t}+A_{+}^{*}e^{-i\omega t}$
into the equations above and obtain
 \begin{eqnarray}
(i\omega-\gamma_{\rm th})Q & = & \gamma_{\rm th}\alpha[a_{0}A_{+}^{*}+a_{0}^{*}(A_{-})]\\
\text{[}-i(\Delta_{e}+\omega)+\kappa/2]A_{-} & = & iGa_{0}Q+i\frac{\Delta_{1}a_{0}}{2}e^{-i\varphi} \\
\text{[}i(\Delta_{e}-\omega)+\kappa/2]A_{+}^{*} & = & -iGa_{0}^{*}Q-i\frac{\Delta_{1}a_{0}^{*}}{2}e^{-i\varphi}
\end{eqnarray}

Under the condition $\omega\ll\Delta_e$, the above equations allow us to calculate the photothermal displacement as

\begin{eqnarray*}
(i\omega-\gamma_{\rm th})Q & = & \gamma_{\rm th}\alpha[\frac{a_{0}(-iGa_{0}^{*}Q-i\frac{\Delta_{1}a_{0}^{*}}{2}e^{-i\varphi})}{i\Delta_{e}+\kappa/2}+\frac{a_{0}^{*}(iGa_{0}Q+i\frac{\Delta_{1}a_{0}}{2}e^{-i\varphi})}{-i\Delta_{e}+\kappa/2}]\\
 & = & \gamma_{\rm th}\alpha\frac{a_{0}(-iGa_{0}^{*}Q-i\frac{\Delta_{1}a_{0}^{*}}{2}e^{-i\varphi})(-i\Delta_{e}+\kappa/2)+a_{0}^{*}(iGa_{0}Q+i\frac{\Delta_{1}a_{0}}{2}e^{-i\varphi})(i\Delta_{e}+\kappa/2)}{\Delta_{e}^{2}+\kappa^{2}/4}\\
 & = & \gamma_{\rm th}\alpha\frac{-\Delta_{e}\Delta_{1}e^{-i\varphi}\left|a_{0}\right|^{2}-2G\left|a_{0}\right|^{2}\Delta_{e}Q}{\Delta_{e}^{2}+\kappa^{2}/4}
\end{eqnarray*}

We obtain the analytical solution of $Q$ as follows
\begin{eqnarray}
Q & = & \frac{-\gamma_{\rm th}\alpha\Delta_{e}\Delta_{1}e^{-i\varphi}\left|a_{0}\right|^{2}}{(i\omega-\gamma_{\rm th})(\Delta_{e}^{2}+\kappa^{2}/4)+2\gamma_{\rm th}\alpha G\left|a_{0}\right|^{2}\Delta_{e}}
\end{eqnarray}

We then calculate the cavity transmission 
\begin{eqnarray*}
t & = & 2[a_{0}A_{+}^{*}+a_{0}^{*}A_{-}]\\
 & = & 2\frac{(i\omega-\gamma_{\rm th})}{\gamma_{\rm th}\alpha}Q\\
 & = & \frac{-2\Delta_{e}\Delta_{1}e^{-i\varphi}\left|a_{0}\right|^{2}}{(\Delta_{e}^{2}+\kappa^{2}/4)+\frac{2\alpha G\left|a_{0}\right|^{2}\Delta_{e}}{(i\omega/\gamma_{\rm th}-1)}}.
\end{eqnarray*}

The cavity transmission can be normalized with the normalization factor of $N = -2\Delta_{e}\Delta_{1}\left|a_{0}\right|^{2}/(\Delta_{e}^{2}+\kappa^{2}/4)$ as
\begin{eqnarray}
t_{\rm norm} & = & \frac{1}{1+\frac{2\alpha G\left|a_{0}\right|^{2}\Delta_{e}}{(i\omega/\gamma_{\rm th}-1)(\Delta_{e}^{2}+\kappa^{2}/4)}} \nonumber\\
 & = & \frac{1}{1+\frac{\zeta}{(i\omega/\gamma_{\rm th}-1)}} \label{eqa: t1}
\end{eqnarray}

This equation shows that the normalized transmission takes the same formula in the cases of both amplitude and phase modulation.

\subsection*{4. Contributions of photothermal effects in the experiment}
As mentioned in the main paper, there are several photothermal effects presenting in our experiment. In this section, we will discuss their contributions to the system response.

We first recall that the cavity has a concave-convex geometry and therefore the beam waist is located in proximity of the front mirror, as shown in Fig 1(b) of the main paper. In other words, the spot size at the back mirror is much larger than that at the front mirror. As the photothermal expansion relaxation rate is inversely proportional to the square of the beam radius (see Section 4.2 of the main paper), the photothermal effects manifested at the front mirror are much more prominent. We can therefore neglect the thermal effects of the back mirror.

Concerning the front mirror, four thermal effects are present: the thermal expansion and thermo-optic effect for either the HR coating or the substrate. Their prominence varies greatly but we need only consider their net effect. We should point out that, since the HR coating of the front mirror faces the outwards side of the cavity, the cavity length is expected to increase as the coating thermally expands. The other three thermal effects (thermo-optic effect of coating and substrate, and expansion of substrate) also tend to increase the optical path length of the cavity. As a result, all thermal effects occurring at the front mirror induce an increase in the effective cavity length, suggesting no obvious competing effect. In our experiment the negative value of effective photothermal coefficient $\sigma$ extracted from data fitting agrees with this analysis.

We repeated the experiment presented in the main paper by placing the HR coating on the inner side of the cavity, like more conventional Fabry-Pérot resonators. Under similar experimental conditions (cavity finesse, intracavity power, etc.), we did not observe any visible photothermal interaction, suggesting that the dominant effect in our system comes from the thermo-optic refractive index change of the front mirror’s substrate. However, we could not present any stronger evidence for a complete dismissal of the other effects.

It is important to note that ultimately it does not matter which effect contributes to the interaction. We can assume that the photothermal effects collectively change the cavity response, and our phenomenological model and results hold valid without concerns on what the source of photothermal interaction is. This is, in fact, one of the advantages of our in-situ characterization scheme. We can measure the effective parameters of the net photothermal effects (rather than an individual thermal effect), which are crucial for the operation and analysis of cavity-based systems.


\end{document}